\documentclass[prb,twocolumn,aps,amsmath,amssymb]{revtex4}
\usepackage[T1]{fontenc}
\usepackage[utf8]{inputenc}
\usepackage{graphicx}
\usepackage{amsmath}
\usepackage{amssymb}
\usepackage{dcolumn}
\usepackage{float}
\usepackage{bm}
\usepackage{amsfonts,times}
\usepackage{xcolor}
\usepackage[breaklinks=true,colorlinks,citecolor=blue,linkcolor=blue,urlcolor=blue]{hyperref}
\DeclareMathAlphabet{\bi}{OML}{cmm}{b}{it}

\def\be{\begin{equation}}
\def\ee{\end{equation}}
\def\bearr{\begin{eqnarray}}
\def\eearr{\end{eqnarray}}

\begin{document}
\title{Properties of an $\alpha$-$T_3$ Aharonov-Bohm quantum ring: Interplay of Rashba spin-orbit coupling and topological defect}
\author{Mijanur Islam}
\author{Saurabh Basu}

\affiliation{Department of Physics, Indian Institute of Technology-Guwahati, Guwahati-781039, India.}
\normalsize

\begin{abstract}
In this paper we investigate the interplay of the Rashba spin-orbit coupling (RSOC) and a topological defect, such as a screw dislocation in an $\alpha$-$T_3$ Aharonov-Bohm quantum ring and scrutinized the effect of an external transverse magnetic field therein. Our study reveals that the energy spectrum follows a parabolic dependence on the Burgers vector associated with the screw dislocation. Moreover, its presence results in an effective flux, encompassing the ramifications due to both the topological flux and that due to the external magnetic field. Furthermore, we observe periodic oscillations in the persistent current in both charge and spin sectors, with a period equal to one flux quantum, which, however suffers a phase shift that is proportional to the dislocation present in the system. Such tunable oscillations of the spin persistent current highlights potential application of our system to be used as spintronic devices. Additionally, we derive and analyse the thermodynamic properties of the ring via obtaining the canonical partition function through Euler-Maclaurin formula. In particular, we compute the thermodynamic potentials, free energies, entropy, and heat capacity and found the latter to yield the expected Dulong-Petit law at large temperatures.
\end{abstract}

\maketitle

\section{Introduction}
In recent years, quantum rings (QRs) have garnered significant attention due to their diverse technological applications, including their use in single-photon emitters, nano-flash memories [\onlinecite{Fomi,Nowo}], photonic detectors [\onlinecite{Fomi, Mich}], and their role as qubits for spintronic quantum computing [\onlinecite{Fomi}]. Moreover, QRs serve as a rich subject for exploring topological effects in condensed matter physics [\onlinecite{Fomi}]. These nanostructures are remarkable due to their non-simply connected topology, which gives rise to intriguing energy structures [\onlinecite{Fuhr}]. These structures differ from most other low-dimensional systems such as quantum dots, quantum wires, and quantum wells. QRs can be categorized into two primary types, one-dimensional (1D) rings with a constant radius [\onlinecite{Cheu,FE,Frus,Lork,Kett,Vief,Sple,Soum,Chap}] and two-dimensional rings (2D) with a variable radius [\onlinecite{Tan,Tan1,Tan2,Bula,Duqu,Bakk,Nowa}]. Of particular note, a special case within the 1D QR family has gained significant recognition in the literature, known as the Aharonov-Bohm (AB) rings [\onlinecite{Citr,Citr1,Beli,Berg}]. Currently, there are numerous works that delve into the properties of the AB rings, employing both theoretical and experimental approaches. For instance, AB rings have been studied in connection with the Aharonov-Casher effect [\onlinecite{Citr,Citr1,Beli,Berg,Ahar}], violations of Lorentz symmetry [\onlinecite{Beli}], mesoscopic decoherence [\onlinecite{Hans}], electromagnetic resonators [\onlinecite{Reul}], and the influence of Rashba spin-orbit interactions [\onlinecite{Berg,Aebe,Shel}].

In the framework of geometric theory of defects [\onlinecite{Kata}], elastic deformations induced by topological defects in continuous media are described using a metric. These theory bears a resemblance to the theory of three-dimensional gravity. Within this geometric formalism, the continuous elastic medium is represented as a Riemann–Cartan manifold, where the curvature and torsion are associated with disclinations and dislocations, respectively, present in the medium. Consequently, the Burgers vector and Frank angle are respectively analogous to torsion and curvature. The impact of topological defects on the quantum dynamics of electrons/holes in a crystal has been explored in various physical scenarios [\onlinecite{Mora,Furt}]. Theoretical descriptions of quantum dynamics within a medium containing dislocations have been undertaken for quite some time. For instance, Kawamura [\onlinecite{Kawa}] and Bausch, Schmitz, and Turski [\onlinecite{Baus,Baus1,Turs}] investigated the scattering of a single particle within dislocated media using a distinct approach. They demonstrated that the equation describing the scattering of a quantum particle by a screw dislocation follows the Aharonov–Bohm form [\onlinecite{Bohm}]. The Aharonov–Bohm effect has also been explored using the Katanaev–Volovich approach in media with a dislocation, as seen in Refs. [\onlinecite{Azev,Azev1}], and in the presence of dislocations, as observed in Refs. [\onlinecite{Furt,Ribe}]. In Ref. [\onlinecite{Aure}], Aurell probed deeper into the influence of dislocations on the properties of quantum dots. More recently, the study of the impact of topological defects in mesoscopic systems has been conducted in Ref. [\onlinecite{Rosa}], particularly with regard to a quantum dot in presence of a dislocation.

Undoubtedly, one of the most promising materials of the century, graphene [\onlinecite{Graph_exp1,Graph_exp2,Graph_exp3,Graph_exp4}], has attracted substantial attention in both theoretical and experimental investigations of quantum ring systems. This heightened interest primarily arises from its remarkable properties. These include the involvement of linearly dispersive 'massless' Dirac fermions [\onlinecite{Zh,De}], the potential emergence of topological phases resulting due to the violation of time reversal symmetry [\onlinecite{Haldane}], and the manifestation of Aharonov-Bohm oscillations in the presence of a magnetic field [\onlinecite{Grap_Lith1,Grap_Lith2,Grap_Lith3,PR,Fa}], among others. Recently, numerous studies have been carried out to unravel the microscopic intricacies of graphene quantum rings under external magnetic fields, both with and without the introduction of spin-orbit interactions [\onlinecite{PR,Graph_Numr3,Graph_Numr4,Graph_Numr5,Graph_Numr6,Graph_Numr7,
Graph_Numr9,Graph_Numr10,Graph_Numr11,Graph_Model1,Graph_Model2,
Graph_Bilayer,hybrid_grapR}]. These investigations have illuminated the potential applications of graphene quantum rings in future optoelectronic [\onlinecite{Graph_Opto}] and interferometric devices [\onlinecite{Graph_Interf}]. 

In recent years, the $\alpha$-$T_3$ system has garnered considerable interest. By adjusting the parameter $\alpha$ in the range of [0:1], the $\alpha$-$T_3$ lattice offers a smooth transition between the honeycomb structure of graphene ($\alpha=0$) and the dice lattice ($\alpha=1$) [\onlinecite{MI,Su,Vi}]. This system can be experimentally realized in heterostructures and optical lattice setups, as previously proposed in various studies [\onlinecite{Wa,Urba,Malc}]. A nearest-neighbor tight-binding analysis reveals that the $\alpha$-$T_3$ lattice accommodates massless quasiparticles that obey the Dirac-Weyl equation, characterized by a generalized pseudospin dependent on the parameter $\alpha$. Numerous investigations have been carried out in recent years to explore a wide array of equilibrium and nonequilibrium properties of the $\alpha$-$T_3$ lattice [\onlinecite{T3_Hall1, T3_Hall2, Klein2, Weiss, ZB, Plasmon1, Plasmon2, Plasmon3, Plasmon4, Mag_Opt1, Mag_Opt2, Mag_Opt3, Mag_Opt4, RKKY1, RKKY2, Min_Con, Ghosh_Topo, spin_hall, aT3_Mijanur}]. On another front, topological phases in the Haldane dice lattice model [\onlinecite{Ghosh_Topo,Sayan_dice}] and the Rashba dice model [\onlinecite{Wa,Rahul1,Rahul2}] have attracted significant attention in recent years. Additionally, the observation of a quantum spin Hall phase transition in the $\alpha$-$T_3$ lattice has been noted [\onlinecite{spin_hall}].

Driven by the promising potential of QRs, this study focuses onto the impact of a topological defect positioned at the center of the $\alpha$-$T_3$ QR, in presence of a Rashba spin-orbit coupling (RSOC). The interplay of RSOC and topological defect on the energy spectra, transport and thermodynamic properties will comprise of the important discussions made in our paper. The introduction of this topological defect is achieved through a geometric description. Furthermore, our analysis involves into the dynamics of a charged particle constrained to navigate a QR with a fixed radius, all while being influenced by the presence of this topological defect. Additionally, we explore the Rashba SOC term, which adheres to the symmetries of the parent $\alpha$-$T_3$ ring. Notably, this term can be manipulated using an external electric field [\onlinecite{Kane}], adding heavy adatoms etc., that effectively breaks the mirror symmetry with respect to the $\alpha$-$T_3$ plane. This discussion underscores the necessity for a comprehensive investigation into the behavior of an $\alpha$-$T_3$ QR with Rashba SOC and topological defect. To further elucidate the controllability of persistent currents, we incorporate an external magnetic field into our analysis. These explorations have sparked substantial research activity, particularly in the context of potential applications within the emerging field of spintronics. An intriguing avenue for further investigation involves the combination of Rashba SOC with $\alpha$-$T_3$ to examine various properties. At the theoretical level, it is imperative to gain a deeper understanding of the cumulative effects of a robust electric field from SOC, confinement, and disorder potentials, such as the topological defect considered by us. These factors directly impact charge transport and the spin-related attributes of electrons, ultimately influencing thermodynamic properties.

Moreover, the exploration of the physical properties of such systems, with a particular focus on their thermal characteristics, holds significant importance in the condensed matter community, particularly in the context of solid-state materials. This endeavour is driven by both practical requirements and the pursuit of fundamental scientific knowledge [\onlinecite{Bala}]. It is noteworthy that previous works [\onlinecite{Alof,Cag,Oliv,Houc}] have explored into the thermal properties of materials like graphene, graphite, carbon nanotubes, and nanostructured carbon materials. However, the thermodynamic properties of an $\alpha$-$T_3$ fermion with Rashba SOC confined in an Aharonov-Bohm (AB) ring and featuring a topological defect have remained unexplored until now. This study also aims to shed light on the thermal properties of an AB $\alpha$-$T_3$ ring. To achieve this, we consider a set of noninteracting, indistinguishable fermions confined within an AB ring. Furthermore, we have observed that the presence of a topological defect results in a non-degenerate energy spectrum. This observation enables us to adopt a strong field approach and utilize a numerical method based on the Euler-MacLaurin formula to compute the canonical partition function. Notably, our results demonstrate the recovery of the well-known Dulong-Petit law for specific heat of solids as a limiting case.

This work is structured as follows. In Section \ref{formalism}, we present the model incorporating a topological defect. Section \ref{results} provides an overview of the electronic properties of the system. The discussion of the thermodynamic properties, including explicit formulas, is presented in various sub-sections of Section \ref{Thermodyna}. Finally, our findings are summarized in Section \ref{Sum}.

\begin{figure}[h!]
\centering
\includegraphics[width=6cm, height=5cm]{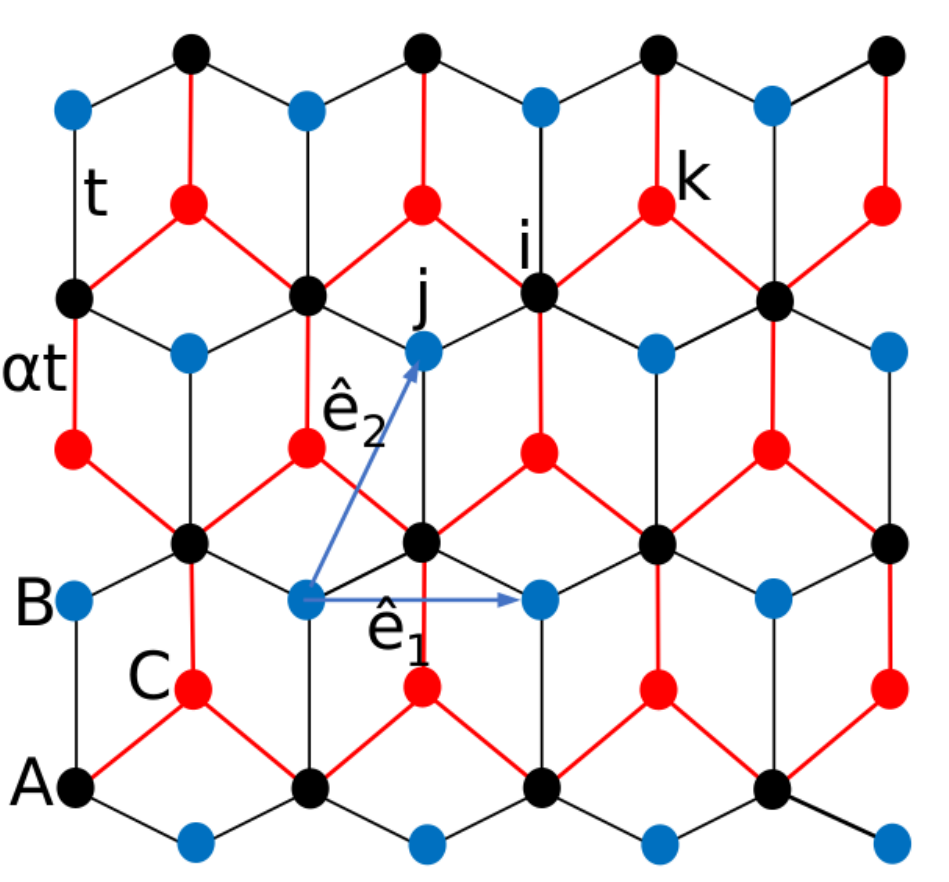}
\caption{(Color online) Lattice structure of the $\alpha$-$T_3$ lattice. Here, A, B, and C lattice sites are shown by black, blue, and red dots respectively. Hopping amplitude between the A and B sublattice is $t$, while between A and C is $\alpha t$. Blue arrows labeled by $\hat{e_1}$ and $\hat{e_2}$ indicate the two translational vectors of the $\alpha$-$T_3$ lattice.}
\label{fig:ring_geo}
\end{figure}

\section{$\alpha$-$T_3$ AB ring with screw dislocation in addition with RSOC}
\label{formalism}
In this section we shall investigate the dynamics of particles in a medium with a screw dislocation along the $z$ direction. The three-dimensional geometry of the medium, in this case, is characterized by non-trivial torsion which is identified with the surface density of the Burgers vector. Thus, the Burgers vector can be viewed as flux of torsion. The screw dislocation is described by the following metric, in cylindrical coordinates,
\begin{equation}
\label{Metric}
ds^2=g_{ij}dx^idy^j=d\rho^2+(dz+\eta d\theta)^2+\rho^2d\theta^2
\end{equation}
with $\rho >0$, $0\leqslant \theta \leqslant 2\pi$, and $-\infty \leqslant z\leqslant \infty$, and where $\eta$ is a parameter related to the Burgers vector $b^z$ by $\eta=b^z/2\pi$.

Further, when we make a measurement of a physical vector quantity, however, we require the components of the vector in the original flat space (the laboratory coordinates). For example, the expectation value of the momentum is obtained by using the momentum operator,
\begin{equation}
\begin{split}
\hat{{\bf p}}=-i\hbar\nabla=-i\hbar \hat{e_i}\zeta^{ij}\partial_j=-i\hbar \hat{e_i}\partial^i,\\
\langle p^i \rangle = \langle \psi|\hat{p}^i|\psi\rangle, i=x,y,z,
\end{split}
\end{equation}
where $\zeta^{ij}=\delta^{ij}$ is the flat-space metric, and $\psi$ is the wave function. If the wave function on the constrained surface is given, we transform the momentum operator as follows [\onlinecite{Ram}],
\begin{equation}
\label{Momentum}
\hat{p}^i=-i\hbar\partial^i=-i\hbar g^{\mu\nu}\frac{\partial x^i}{\partial \tilde{x}^\mu}\frac{\partial}{\partial\tilde{x}^\nu}=-i\hbar\tilde{\partial^\mu}x^i\tilde{\partial_\mu}
\end{equation}
The coordinate indices in the curved space are given by Greek letters, and the curved space coordinates are denoted by $\tilde{x}^\mu$, with $\tilde{\partial_\mu}$ being the covariant derivative in the curved space. Concisely, the momentum operators for the metric given in Eq. (\ref{Metric}) can be written in vector form as,
\begin{equation}
\label{Momentum1}
\begin{split}
p^x &= -i\hbar\Big[\cos\theta\frac{\partial}{\partial \rho}-\frac{\sin\theta}{\rho}\frac{\partial}{\partial\theta}+\frac{\eta\sin\theta}{\rho}\frac{\partial}{\partial z} \Big],\\
p^y &= -i\hbar\Big[\sin\theta\frac{\partial}{\partial \rho}+\frac{\cos\theta}{\rho}\frac{\partial}{\partial\theta}-\frac{\eta\cos\theta}{\rho}\frac{\partial}{\partial z} \Big],\\
p^z &= -i\hbar\Big[\frac{\eta^2+\rho^2}{\rho^2}\frac{\partial}{\partial z}-\frac{\eta}{\rho^2}\frac{\partial}{\partial \theta} \Big].
\end{split}
\end{equation}
We shall use these momenta to study the Rashba spin-orbit coupled Aharonov-Bohm $\alpha$-$T_3$ ring in presence of a topological defect.

Now, we consider the $\alpha$-$T_3$ quantum ring system includes Rashba SOC. The corresponding Hamiltonian can be written as, $H=H_0+H_R$, where $H_0$ is the tight-binding term, and $H_R$ is the Rashba spin-orbit coupling term. We write the Hamiltonian as [\onlinecite{Rash}],
%\begin{widetext}
\begin{equation}
\label{Ham_Rash}
\begin{aligned}
H &=-t\sum_{\langle ij\rangle\sigma}c_{i\sigma}^\dagger c_{j\sigma}-\alpha t\sum_{\langle ik\rangle\sigma}c_{i\sigma}^\dagger c_{k\sigma}\\
&-i\lambda_R\sum_{\langle ij\rangle\sigma\sigma^\prime}c_{i\sigma}^\dagger(\hat{D}_{ij}\cdot\vec{\tau})_{\sigma\sigma^\prime}c_{j\sigma^\prime}\\
&-i\alpha\lambda_R\sum_{\langle ik\rangle\sigma\sigma^\prime}c_{i\sigma}^\dagger(\hat{D}_{ik}\cdot\vec{\tau})_{\sigma\sigma^\prime}c_{k\sigma^\prime}+\bigg (h.c.\bigg),
\end{aligned}
\end{equation}
%\end{widetext}
where $\sigma = \uparrow, \downarrow$, spin indices and $i,j,k$ are labels for the sites corresponding to A, B, and C sublattices respectively. The first term is the electron hopping between the A and B sites, while the second one is that between the A and C sites. The summation of $\langle ij \rangle$ ($\langle ik \rangle$) runs over the nearest neighbour sites of A-B (A-C). Further, the Rashba SOC induced by electric fields due to a gradient of the crystal potential [\onlinecite{Wa,Rahul1,Rahul2}], where $\vec{\tau}=(\tau_x,\tau_y,\tau_z)$ is the Pauli matrix vector, $\hat{D}_{ij}$ ($\hat{D}_{ik}$) is the unit vector along the direction of the cross product $\vec{E}_{ij} \times \vec{r}_{ij}$ ($\vec{E}_{ik} \times \vec{r}_{ik}$) of the electric field $\vec{E}_{ij}$ ($\vec{E}_{ik}$) and displacement $\vec{r}_{ij}$ ($\vec{r}_{ik}$) for the bond $ij$ ($ik$). $\lambda_R$ is the strength of Rashba SOC between the A and the B sites, while $\alpha\lambda_R$ is that between the A and the C sites. In momentum space, the Hamiltonian of the $\alpha$-$T_3$ lattice becomes,
\begin{widetext}
\begin{equation}
\label{Ham_Rash1}
H=\begin{pmatrix}
0 & -t\gamma_\mathrm{k}^* & 0 & 0 & -i\lambda_R\gamma_{\mathrm{k}+}^* & 0\\
-t\gamma_\mathrm{k} & 0 & -\alpha t\gamma_\mathrm{k}^* & i\lambda_R\gamma_{\mathrm{k}-} & 0 & i\alpha\lambda_R\gamma_{\mathrm{k}+}^*\\
0 & -\alpha t\gamma_\mathrm{k} & 0 & 0 & -i\alpha\lambda_R\gamma_{\mathrm{k}-} & 0\\
0 & -i\lambda_R\gamma_{\mathrm{k}-}^* & 0 & 0 & -t\gamma_\mathrm{k}^* & 0\\
i\lambda_R\gamma_{\mathrm{k}+} & 0 & i\alpha\lambda_R\gamma_{\mathrm{k}-}^* & -t\gamma_\mathrm{k} & 0 & -\alpha t\gamma_\mathrm{k}^*\\
0 & -i\alpha\lambda_R\gamma_{\mathrm{k}+} & 0 & 0 & -\alpha t\gamma_\mathrm{k} & 0
\end{pmatrix}
\end{equation}
\end{widetext}
we defined $\gamma_\mathrm{k}=1+e^{ik_1}+e^{ik_2}$ and $\gamma_{\mathrm{k}\pm}=1+e^{i(k_1\pm 2\pi/3)}+e^{i(k_2\pm 4\pi/3)}$, where the components are along the axes indicated in Fig. \ref{fig:ring_geo} as $k_i=\vec{k}\cdot\hat{\mathrm{e}}_i$. Our basis is $(c_{1\mathrm{k}\uparrow},c_{2\mathrm{k}\uparrow},c_{3\mathrm{k}\uparrow},c_{1\mathrm{k}\downarrow},c_{2\mathrm{k}\downarrow},c_{3\mathrm{k}\downarrow})$.

 In the vicinity of a Dirac point (namely, {\bf K}), and taking the momentum correction (\ref{Momentum1}) due to screw dislocation into account, the Hamiltonian (\ref{Ham_Rash1}) corresponding to an ideal $\alpha$-$T_3$ ring is given by  [\onlinecite{FE,Spin_dep1,Spin_dep2,Bol,DR,Graph_Model1, Graph_Model2,Mijanur}],
\begin{widetext}
\begin{equation}
\resizebox{\hsize}{!}{$
H_{ring}=\frac{\hbar v_F}{R}\begin{pmatrix}
0 & -i(m-k\eta+\frac{1}{2})\cos\xi e^{\frac{i\pi}{3}} & 0 & 0 & -\frac{\lambda_R}{t}(m-k\eta+\frac{1}{2})\cos\xi e^{\frac{i\pi}{3}} & 0\\
i(m-k\eta+\frac{1}{2})\cos\xi e^{-\frac{i\pi}{3}} & 0 & -i(m-k\eta-\frac{1}{2})\sin\xi e^{\frac{i\pi}{3}} & \frac{\lambda_R}{t}(m-k\eta-\frac{1}{2})\cos\xi e^{\frac{i\pi}{3}} & 0 & -\frac{\lambda_R}{t}(m-k\eta+\frac{1}{2})\sin\xi e^{-\frac{i\pi}{3}}\\
0 & i(m-k\eta-\frac{1}{2})\sin\xi e^{-\frac{i\pi}{3}} & 0 & 0 & -\frac{\lambda_R}{t}(m-k\eta-\frac{1}{2})\sin\xi e^{-\frac{i\pi}{3}} & 0\\
0 & \frac{\lambda_R}{t}(m-k\eta-\frac{1}{2})\cos\xi e^{-\frac{i\pi}{3}} & 0 & 0 & i(m-k\eta-\frac{1}{2})\cos\xi e^{-\frac{i\pi}{3}} & 0\\
-\frac{\lambda_R}{t}(m-k\eta+\frac{1}{2})\cos\xi e^{-\frac{i\pi}{3}} & 0 & \frac{\lambda_R}{t}(m-k\eta-\frac{1}{2})\sin\xi e^{\frac{i\pi}{3}} & -i(m-k\eta-\frac{1}{2})\cos\xi e^{\frac{i\pi}{3}} & 0 & i(m-k\eta+\frac{1}{2})\sin\phi e^{-\frac{i\pi}{3}}\\
0 & -\frac{\lambda_R}{t}(m-k\eta+\frac{1}{2})\sin\xi e^{\frac{i\pi}{3}} & 0 & 0 & -i(m-k\eta+\frac{1}{2})\sin\xi e^{\frac{i\pi}{3}} & 0
\end{pmatrix}$}
\end{equation}
\end{widetext}
where $\tan\xi=\alpha$ and $\hbar v_F=3at/2\cos\xi$. The eigenstates of the ring Hamiltonian can be obtained as,
\begin{equation}
\label{Wave}
\psi(R,\theta)=\begin{pmatrix}
\chi_{1\uparrow}(R)e^{i\theta}\\
\chi_{2\uparrow}(R)\\
\chi_{3\uparrow}(R)e^{-i\theta}\\
\chi_{1\downarrow}(R)e^{-i\theta}\\
\chi_{2\downarrow}(R)\\
\chi_{3\downarrow}(R)e^{i\theta}
\end{pmatrix}e^{im\theta}e^{ikz}
\end{equation}
where the integer $m$ labels the orbital angular momentum quantum number, $k$ is the momentum along the $z$ direction and $\chi_i$'s denote the amplitudes corresponding to the three sublattices. Here, we investigate the electronic behaviour at a given value of radius $\rho$, namely $\rho = R$, such that the radial part is rendered frozen in the eigensolutions. For the sake of the hermiticity of the Hamiltonian in ring geometry, we made the replacements $\rho \to R$ and $\frac{\partial}{\partial \rho} \to -\frac{1}{2R}$ and obtain the energy spectrum as,
\begin{equation}
\label{Egn_Rash}
\begin{aligned}
E_{1} &= 0\\
E_{2} &= \kappa\frac{\epsilon}{2}\Bigg\{\bigg[1+4\big(m-k\eta\big)^2-4\big(m-k\eta\big)\frac{1-\alpha^2}{1+\alpha^2}\bigg]\\
& \bigg(1+\frac{\lambda_R^2}{t^2}\bigg)\Bigg\}^\frac{1}{2}\\
E_{3} &= \kappa \frac{\epsilon}{2}\Bigg\{\Big[1+4\big(m-k\eta\big)^2\Big]\bigg(1+\frac{\lambda_R^2}{t^2}\frac{1-\alpha^2}{1+\alpha^2}\bigg)+\\
& 4\big(m-k\eta\big)\bigg(\frac{\lambda_R^2}{t^2}+\frac{1-\alpha^2}{1+\alpha^2}\bigg)\Bigg\}^\frac{1}{2}
\end{aligned}
\end{equation}
where $\kappa=\pm 1$ is the particle-hole index and $\epsilon=\frac{\hbar v_F}{R}$. $E_1$ is the zero energy flat band, $E_2$ and $E_3$ correspond to the energies for the $\uparrow$-spin and $\downarrow$-spin bands respectively. The energy spectra at the {\bf K}-valley in presence of Rashba SOC  and screw dislocation are expressed in Eq. (\ref{Egn_Rash}).

\section{Results and discussions}
\label{results}
\subsection{No magnetic field}
\begin{figure}[h!]
\centering
\includegraphics[width=9cm, height=10cm]{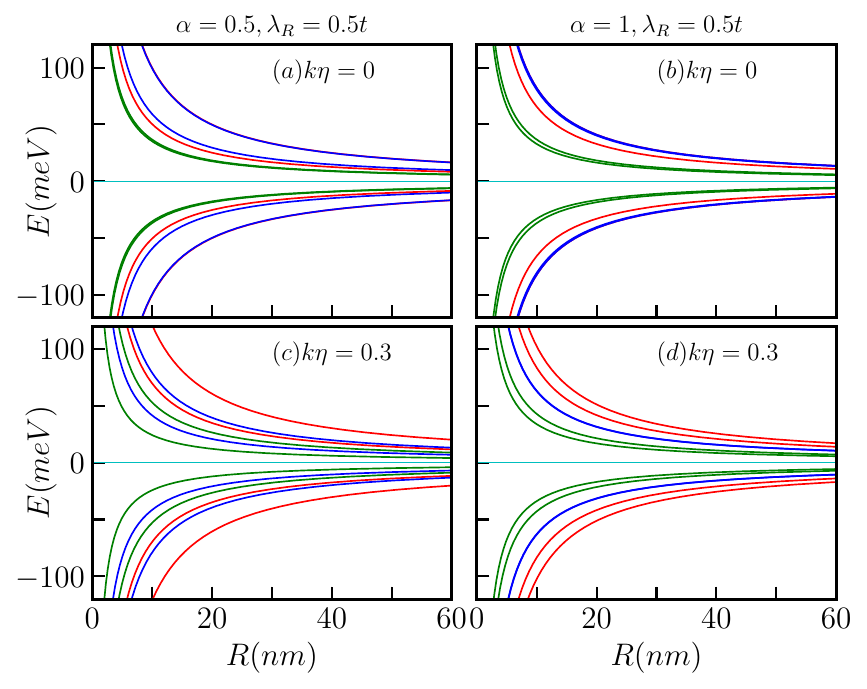}
\caption{(Color online) The spin-split energy spectra in the absence of a magnetic field for the $\alpha$-$T_3$ AB ring are depicted as a function of the ring radius,  $R$, for quantum numbers $m=0$ (green curves), $m=1$ (blue curves), and $m=-1$ (red curves). Results in the absence of a topological defect are presented in panels (a) for $\alpha=0.5$ and (b) for $\alpha=1$. Panels (c) and (d) showcase the effects of a topological defect with a magnitude of $k\eta=0.3$ for $\alpha=0.5$ and $\alpha=1$ respectively. The Rashba coupling parameter is all the while set at $\lambda_R=0.5t$.}
\label{fig:EvsR}
\end{figure}
{\it Without screw dislocation ($k\eta=0$)}: We exclude the effects of topological defect, or what we call as screw discolation ($k\eta=0$) to begin with. Fig. \ref{fig:EvsR}(a) and (b) display the energies as a function of the ring radius, $R$, for various values of $\alpha$ at a fixed $\lambda_R$ value. One can easily verify the results of the $\alpha$-$T_3$ quantum ring without the RSOC term [\onlinecite{Mijanur}] by setting $\lambda_R = 0$ ($k\eta=0$ anyway) in Eq. (\ref{Egn_Rash}). In this case, we have considered a particular value for the RSOC, namely, $\lambda_R = 0.5t$ corresponding to $\alpha=0.5$ and $1$, and plotted only the $m = -1$, $0$, and $1$ bands represented by red, green, and blue curves, respectively. When $\lambda_R=0$, the system exhibits three bands, with one being a flat band. However, with a non-zero $\lambda_R$ the original three bands split into six spin dependent bands, including two non-dispersive flat bands and four dispersive bands as described by Eq. (\ref{Egn_Rash}). From Fig. \ref{fig:EvsR}, it is evident that all the energy branches have a 1/$R$ dependence and approach $E\to 0$ for very large radii, irrespective of the value of $\alpha$. Additionally, the dispersive bands remain non-degenerate, in contrast to the case of the pseudospin-1 $\alpha$-$T_3$ QR without SOC. Moreover, the dispersive $\uparrow$-spin and $\downarrow$-spin bands split as well. Specifically, for $m=0$, the energies are given by,
\begin{equation}
E_2=\frac{\kappa \epsilon}{2}\sqrt{1+\frac{\lambda_R^2}{t^2}}
\end{equation}
and 
\begin{equation}
E_{3}=\frac{\kappa\epsilon}{2}\sqrt{1+\frac{\lambda_R^2}{t^2}\frac{1-\alpha^2}{1+\alpha^2}}.
\end{equation}
It can be observed that the $\uparrow$-spin energy band, $E_{2}$ is independent of $\alpha$, while  the $\downarrow$-spin band, $E_{3}$ has a dependency on $\alpha$. This is an interesting result, since the $\uparrow$-spin band energies are insensitive to whether we are talking about graphene or the dice lattice. Consequently, the splitting between the $m=0$ bands (green curves in Fig. \ref{fig:EvsR}) increases with increasing values of $\alpha$. Whereas, the splitting between the bands with $m=-1$ (red curves in Fig. \ref{fig:EvsR}) and $m=1$ (blue curves in Fig. \ref{fig:EvsR}) decreases as $\alpha$ increases. Furthermore, the energy splitting decreases with increase of $|m|$ values for all values of $\alpha$. In addition to that, the energy splitting increases with an increase of $\lambda_R$. An intriguing observation is that for $\alpha=1$, i.e., the dice lattice, the energies are given by,
\begin{equation}
E_{2}=\frac{\kappa\epsilon}{2}\sqrt{(1+4m^2)(1+\frac{\lambda_R^2}{t^2})}
\end{equation}
and
\begin{equation}
E_{3}=\frac{\kappa\epsilon}{2}\sqrt{1+4m^2+4m\frac{\lambda_R^2}{t^2}}.
\end{equation}
Thus, $E_{2}$ is a even function of $m$, making it two-fold degenerate corresponding to $m=\pm1,\pm2, \pm3,....$ etc. On the other hand, the $E_{3}$ band is an odd function of $m$, resulting in it being non-degenerate as illustrated in Figs. \ref{fig:EvsR}(b).

\begin{figure}[h!]
\centering
\includegraphics[width=9cm, height=8cm]{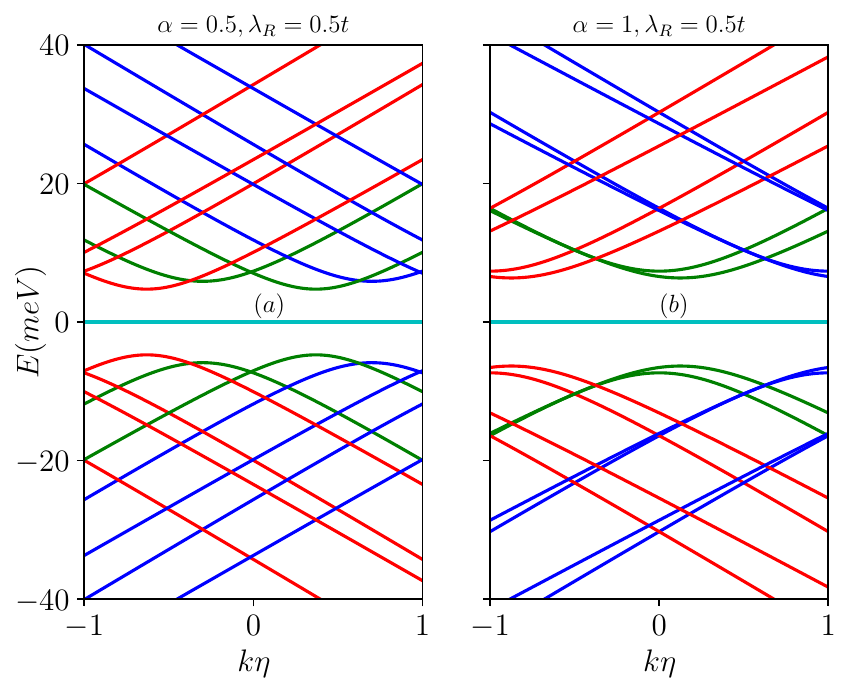}
\caption{(Color online) Energy levels in the absence of a magnetic field, as a function of the screw dislocation parameter $k\eta$, are depicted for (a) $\alpha=0.5$ and (d) $\alpha=1$, with a fixed ring radius of $R=50$ $nm$. The energy levels corresponding to different total angular momentum quantum numbers, namely $m=-1,-2$ (red curves), $m=0$ (green curves), and $m=1,2$ (blue curves), are presented. The Rashba coupling parameter remains constant at $\lambda_R=0.5t$.}
\label{fig:Evsdis}
\end{figure}
{\it With screw dislocation ($k\eta\neq 0$)}: The scenario involving a topological defect arises when we implement the transformation via $m^\prime \to (m-k\eta)$ in the results of the Euclidean model. Consequently, upon examining Eq. (\ref{Egn_Rash}), it becomes apparent that, unlike the situation where $k\eta=0$, all the energy levels for both $\uparrow$-spin and $\downarrow$-spin states, including the $m = 0$ bands, become dependent on the parameters $\alpha$ and $\lambda_R$. Hence, the splitting between the spin-split bands hinges on the presence of dislocations in the system, the parameter $\alpha$, the strength of the Rashba coupling, $\lambda_R$, as dictated by Eq. (\ref{Egn_Rash}). Furthermore, for $\alpha = 1$, it is worth noting that $E_2$ and $E_3$ are non-degenerate, in contrast to the previous case. These findings are illustrated in Figs. \ref{fig:EvsR}(c) and (d). Additionally, the quantum number $m$ undergoes a shift that depends on $\eta$, which is associated with the Burgers vector. This shift is a manifestation of the Aharonov-Bohm effect, akin to what is observed in the context of a one-dimensional quantum ring penetrated by a magnetic flux [\onlinecite{Fuhr}]. The energy levels as a function of the strength of the screw dislocation are presented in Fig. \ref{fig:Evsdis} for two distinct cases, namely, $\alpha = 0.5$ and $\alpha = 1$ (dice lattice). Moreover, Eq. (\ref{Egn_Rash}) makes it clear that the dispersive energy levels exhibit a hyperbolic dependence on the screw dislocation. It is important to note that the energy spectrum depends on the Burgers vector, $b^z=2\pi \eta$, as well as the radius, $R$ of the ring, and hence associated with the geometry of the system.

Additionally, there are energy extrema, with maxima in the valence band and minima in the conduction band, that pertain to different spin bands. The positions of these extrema are determined by the strength of dislocation as follows. For the $\uparrow$-spin bands, the extrema are found at
\begin{equation}
k\eta=m-\frac{1}{2}\frac{1-\alpha^2}{1+\alpha^2}
\end{equation}
which is independent of $\lambda_R$. For the dice case ($\alpha=1$), the extrema occur at $k\eta=m$. However, for the $\downarrow$-spin bands, the extrema occurs at
\begin{equation}
k\eta=m+\frac{1}{2}\frac{\frac{\lambda_R^2}{t^2}+\frac{1-\alpha ^2}{1+\alpha^2}}{1+\frac{\lambda_R^2}{t^2}\frac{1-\alpha^2}{1+\alpha^2}}.
\end{equation}
The expression above shows a dependency on $\lambda_R$ and hence we shall observe an interplay between the dislocation and RSOC. In fact, this interplay gives rise to interesting consequences as we shall see later. Furthermore, for each value of $m$, a band crossing point exists for $\alpha < 1$. However, in the case of $\alpha = 1$, no band crossings occur, instead, the bands touch each other at certain specific $k\eta$ values, depending upon the parameter $\lambda_R$.

\begin{figure}[h!]
\centering
\includegraphics[width=9cm, height=10cm]{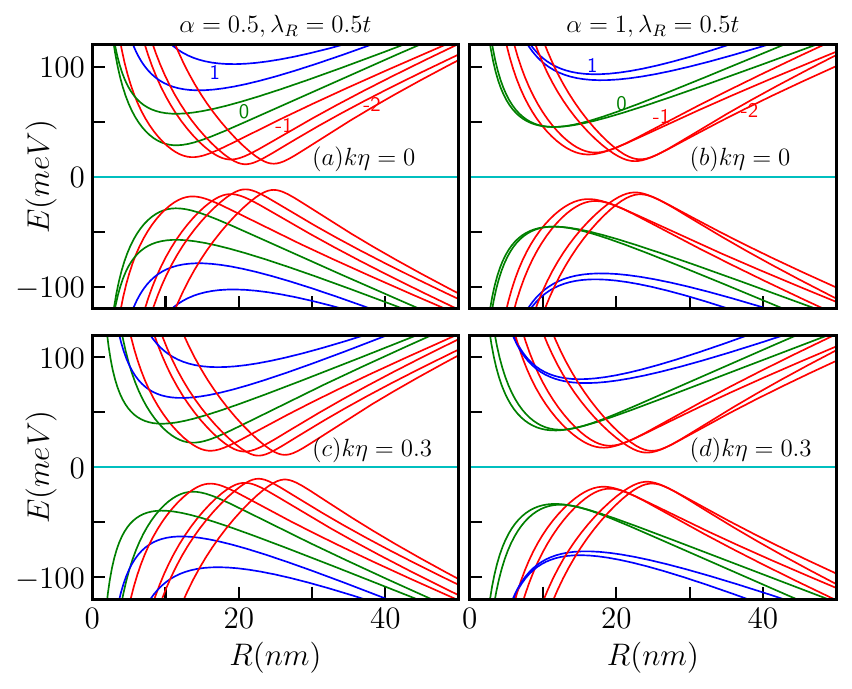}
\caption{(Color online) The energy spectra as a function of the ring radius $R$ of the $\alpha$-$T_3$ quantum ring under the influence of an external magnetic field with a strength of $B_0=5$T at the ${\bf K}$-valley are presented in panels (a) for $\alpha=0.5$ and (b) for $\alpha=1$, without topological defect. In panels (c) and (d), we introduce a screw dislocation with a magnitude of $k\eta=0.3$ for $\alpha=0.5$ and $\alpha=1$ respectively. The parameters used in these calculations are $\lambda_R=0.5t$ and $t=1 eV$. The quantum number $m$ ranges from $-2$ to $2$, with positive values of $m$ represented by blue curves, $m=0$ depicted by green curves, and negative values by red curves.}
\label{fig:EvsR_mag}
\end{figure}
\subsection{Effect of magnetic field}
Now let us discuss the case when the $\alpha$-$T_3$ ring is threaded by a perpendicular magnetic field ${\bf B}=B_0\hat{z}$. The only non-zero component of the vector potential is $A_\theta=B\rho/2$. Notice that, in the non-Euclidean metric of the dislocation, the vector potential that produces the uniform magnetic field, is identical to the flat space (Euclidean) potential vector. The spectrum of the system is modified by the field flux as follows,
\begin{equation}
\begin{aligned}
E_{1}(\Phi) &= 0\\
E_{2}(\Phi) &= \kappa\frac{\epsilon}{2}\biggl\{\bigg[1+4\big(m-k\eta+\frac{\Phi}{\Phi_0}\big)^2-4\big(m-k\eta+\frac{\Phi}{\Phi_0}\big)\\
&\frac{1-\alpha^2}{1+\alpha^2}\bigg]\bigg(1+\frac{\lambda_R^2}{t^2}\bigg)\biggl\}^\frac{1}{2}\\
E_{3}(\Phi) &= \kappa \frac{\epsilon}{2}\biggl\{\Big[1+4\big(m-k\eta+\frac{\Phi}{\Phi_0}\big)^2\Big]\bigg(1+\frac{\lambda_R^2}{t^2}\frac{1-\alpha^2}{1+\alpha^2}\bigg)+\\
& 4\big(m-k\eta+\frac{\Phi}{\Phi_0}\big)\bigg(\frac{\lambda_R^2}{t^2}+\frac{1-\alpha^2}{1+\alpha^2}\bigg)\biggl\}^\frac{1}{2}
\end{aligned}
\label{Egn_Rash_flux}
\end{equation}
where $\beta=\Phi/\Phi_0$ with $\Phi=\pi R^2B_0$ is magnetic flux through the ring and $\Phi_0$ is the usual flux quantum ($=h/e$). The Zeeman coupling has been neglected at small enough values of the field. The addition of a magnetic field, represented by a U(1) minimal coupling with flux $\Phi$ threading the ring, breaks the time reversal symmetry allowing for the emergence of persistent charge currents [\onlinecite{Butt}] which we shall discuss later.

{\it Without screw dislocation ($k\eta= 0$)}: In Fig. \ref{fig:EvsR_mag}, we show the dependence of a few energy levels on the ring radius, $R$, considering $B_0 = 5$T for the two aforementioned cases i.e., $\alpha=0.5$ and $1$ with $\lambda_R=0.5t$. Each level exhibits a non-monotonic behaviour as a function of the radius $R$. The energy levels attain an extremum (minimum for conduction band and maximum for valence band) at a particular value of $R$. However, the positions of these extrema depend on the values of $m$, $\alpha$ and $\lambda_R$ explicitly. In the limit of small $R$, all the energy levels vary inversely with $R$. On the other hand, the energy scales as, $E \sim |R|$ in limit of large $R$. Additionally, for a fixed magnetic field and for large $m$, the locations of the extrema points for different $m$ depend on $R$ as $R\propto \sqrt{|m|}$ irrespective of $\alpha$. Thus, the concept of large radii differs for different values of $m$. Consequently, for negative values of $m$, the extrema points of the energy exhibit a scaling behaviour, namely, $E_{min}\propto 1/\sqrt{|m|}$, resulting in a diminishing of the spectral gap with increasing $|m|$. Conversely, for positive values of $m$, the energy extrema scales as, $E_{min}\propto \sqrt{m}$. Furthermore, from Eq. (\ref{Egn_Rash_flux}), it is evident that in presence of a magnetic field, the energy splitting between the bands of the $m = 0$ level as well as the $m \neq 0$ levels decreases with the increase in the values of the parameters $\alpha$ and $\lambda_R$. Again, from Eq. (\ref{Egn_Rash_flux}) it is noted that for $\alpha < 1$, there are two points where the spin bands cross each other as a function of $R$ for $m=0$ and negative values of $m$, i.e., $m=-1,-2,-3,...$ etc. bands, whereas there is only one band crossing point for the possitive values, namely, $m=1,2,3...$. These crossings obey the following condition,
\begin{widetext}
\begin{equation}
\label{Eqn_flux}
\frac{\lambda_R^2}{t^2}\Big[1+4\big(m+\frac{\Phi}{\Phi_0}\big)^2-4\big(m+\frac{\Phi}{\Phi_0}\big)\Big]-\frac{\lambda_R^2}{t^2}\frac{1-\alpha^2}{1+\alpha^2}\Big[1+4\big(m+\frac{\Phi}{\Phi_0}\big)^2+4\big(m+\frac{\Phi}{\Phi_0}\big)\Big]-8\big(m+\frac{\Phi}{\Phi_0}\big)\frac{1-\alpha^2}{1+\alpha^2}=0.
\end{equation}
\end{widetext}
Eq. (\ref{Eqn_flux}) can be checked against the plot shown in Fig. \ref{fig:EvsR_mag}(a). Now, for the dice lattice case ($\alpha=1$), the above mentioned condition requires $m+\frac{\Phi}{\Phi_0}=\frac{1}{2}$, which implies that along the radius $R$, the energy band crossing points occur at $R=\sqrt{2}l_0\sqrt{(1/2-m)}$, where $l_0=\sqrt{\hbar/(eB_0)}$ is the magnetic length. Consequently, there is only one band crossing point for $m=0$ and $m=-1,-2,-3,..$ etc. bands. Furthermore, band crossing is prohibited for positive values of $m$ as there are no real values of $R$ for $m>0$, indicating that the corresponding spin bands do not cross each other and hence there will not be any degeneracy as illustrated in Fig. \ref{fig:EvsR_mag}(b) by the blue curves. 

\begin{figure}[h!]
\centering
\includegraphics[width=9cm, height=10cm]{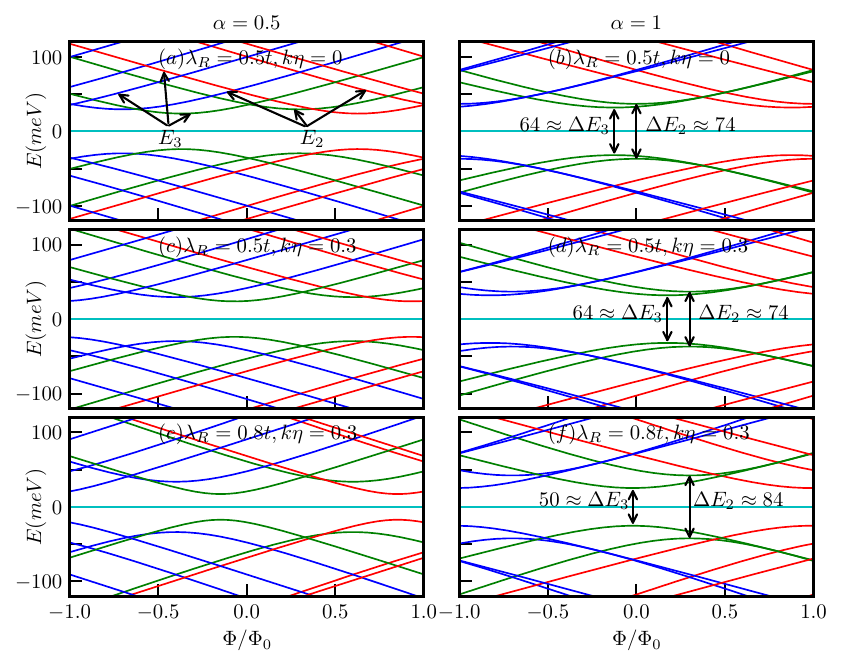}
\caption{(Color online) The energy levels, as functions of the external magnetic flux $\Phi/\Phi_0$ with a fixed ring radius of $R=10$ $nm$, maintain the same conventions for the quantum number $m$ as in the previous plots. We explore various combinations of $\alpha$, $\lambda_R$, and $k\eta$ as follows: (a) $\alpha=0.5$, $\lambda_R=0.5t$ and $k\eta=0$, (b) $\alpha=1$, $\lambda_R=0.5t$ and $k\eta=0$, (c) $\alpha=0.5$, $\lambda_R=0.5t$ and $k\eta=0.3$, (d) $\alpha=1$, $\lambda_R=0.5t$ and $k\eta=0.3$, (e) $\alpha=0.5$, $\lambda_R=0.8t$ and $k\eta=0.3$, and (f) $\alpha=1$, $\lambda_R=0.8t$ and $k\eta=0.3$.}
\label{fig:Evsphi}
\end{figure}

The energy levels as a function of the external magnetic flux ($\Phi/\Phi_0$) are depicted in Figs. \ref{fig:Evsphi}(a) and (b) for a quantum ring with $R=10$ $nm$, considering two cases, namely, $\alpha=0.5$ and $\alpha=1$ with $\lambda_R=0.5$ in unit of $t$. The curves are represented by red, green, and blue colors corresponding to $m=-1$, $m=0$, and $m=1$, respectively. The magnetic field dependence of the energy spectra becomes evident when we rewrite Eq. (\ref{Egn_Rash_flux}) as, 
\begin{equation}
\label{hyper1}
E^2_{2}-\frac{\epsilon^2}{4}\Big[1+4\big(m+\frac{\Phi}{\Phi_0}\big)^2-4\big(m+\frac{\Phi}{\Phi_0}\big)\frac{1-\alpha^2}{1+\alpha^2}\Big]\Big(1+\frac{\lambda_R^2}{t^2}\Big)=0
\end{equation}
and
\begin{equation}
\label{hyper2}
\begin{aligned}
E^2_{3}-\frac{\epsilon^2}{4}\Big[\Big\{1+4\big(m+\beta\big)^2\Big\}\Big(1+\frac{\lambda_R^2}{t^2}\frac{1-\alpha^2}{1+\alpha^2}\Big)+\\
4\big(m+\beta\big)\Big(\frac{\lambda_R^2}{t^2}+\frac{1-\alpha^2}{1+\alpha^2}\Big)\Big]=0.
\end{aligned}
\end{equation} 
Thus, the energies display a hyperbolic dependence on the applied magnetic field, exhibiting extrema at the flux values given by,
\begin{equation}
\label{cond1}
\frac{\Phi}{\Phi_0}=-m+\frac{1}{2}\frac{1-\alpha^2}{1+\alpha^2}
\end{equation}
for $\uparrow$-spin band $E_{2}$, the extrema points are independent of the strength of the Rashba coupling, but depends on the values of $m$ and the parameter $\alpha$. For the dice lattice ($\alpha=1$), the extrema occur at $\Phi/\Phi_0=-m$. However, the extrema for the $\downarrow$-spin band $E_{3}$ occur at
\begin{equation}
\label{cond2}
\frac{\Phi}{\Phi_0}=-m-\frac{1}{2}\frac{\frac{\lambda_R^2}{t^2}+\frac{1-\alpha ^2}{1+\alpha^2}}{1+\frac{\lambda_R^2}{t^2}\frac{1-\alpha^2}{1+\alpha^2}},
\end{equation}
showing a dependency on the strength of Rashba SOC, $\alpha$ and $m$. For the dice lattice, the extrema are obtained at $\Phi/\Phi_0=-m-\frac{1}{2}\frac{\lambda_R^2}{t^2}$. The energy gaps at the extrema points, that is, the minimum values of the gap are given by,
\begin{equation}
\label{gap}
\begin{aligned}
\Delta E_{2} &=\frac{2\epsilon\alpha}{1+\alpha^2}\sqrt{1+\frac{\lambda_R^2}{t^2}},\\
\Delta E_{3} &=\frac{2\epsilon\alpha}{1+\alpha^2}\sqrt{\frac{1-\frac{\lambda_R^4}{t^4}}{1+\frac{\lambda_R^2}{t^2}\frac{1-\alpha^2}{1+\alpha^2}}}.
\end{aligned}
\end{equation}
 Therefore, it is observed that for a fixed value of Rashba coupling, the energy gaps for both the spin bands increase with increase in $\alpha$. However, the minimum energy gap for both the spin bands is independent of $m$. Also the $\downarrow$-spin bands, namely, $E_{3}$ have lower energies than the $\uparrow$-spin bands ($E_2$). The $\uparrow$-spin $E_2$ and $\downarrow$-spin $E_3$ bands are illustrated in the Fig. \ref{fig:Evsphi}(a). For the dice lattice case, and for $\lambda_R=0.5t$, the energy gaps are obtained as, $\Delta E_{2}\approx 74$ $meV$ and $\Delta E_{3}\approx 64$ $meV$ which can be verified from the Fig. \ref{fig:Evsphi}(b). Furthermore, from Fig. \ref{fig:Evsphi} it is evident that $E_{2}(m)\neq E_{2}(-m)$ and $E_{3}(m)\neq E_{3}(-m)$, indicating the existence of finite spin currents.

{\it With screw dislocation ($k\eta\neq 0$)}: Let us now delve into the impact of a topological defect, specifically a screw dislocation, on the electronic spectra of the $\alpha$-$T_3$ QR. Referring to Equation (\ref{Egn_Rash_flux}), we find that all of the previously discussed characteristics remain unaltered, except we may consider that $m$ as an effective quantum number, let us call it as $m^\prime$ with $m^\prime \to (m - k\eta)$ which has a shift in the $m$ values by an amount of $k\eta$. The results are visually represented in Figs. \ref{fig:EvsR_mag}(c) and (d), while considering $\alpha$ values of $0.5$ and $1$, with the parameter $k\eta$ set to a modest value, say $k\eta=0.3$. Again, for the case of $\alpha = 1$, the band crossing points emerge at $R=\sqrt{2}l_0\sqrt{\{1/2-(m-k\eta)\}}$, signifying a shift to the right as depicted in Fig. \ref{fig:EvsR_mag}(d).

Furthermore, the conditions for the extremal points defined in Eqs. (\ref{cond1}) and (\ref{cond2}) are also modified due to $m^\prime \to (m - k\eta)$, resulting in a displacement of these extremal points by an amount equal to $k\eta$ in the positive direction, as illustrated in Figs. \ref{fig:Evsphi}(c)-(f). It is important to note that the energy gaps at these extremal points, as per Eq. \ref{gap}, are not influenced by the presence of the topological defect ($k\eta$), their dependencies solely rely on the parameters $\alpha$ and $\lambda_R$. To demonstrate this, we have considered two values of $\lambda_R$, namely, $\lambda_R = 0.5t$ and $\lambda_R = 0.8t$. The energy gap values for the $\alpha = 1 $ case are presented in the respective figures (see Fig. \ref{fig:Evsphi}).

In this particular scenario, we are dealing with a one-dimensional quantum ring in the presence of a screw dislocation and Aharonov–Bohm flux. The energy spectrum displays a parabolic dependence on the Burgers vector, similar to its dependence on the magnetic flux. Upon analysing our findings, we can infer that a particle within a space featuring a topological defect behaves similarly to a particle in a Euclidean space in the presence of an effective magnetic flux traversing the ring. This effective flux is a combination of two contributions, the first is of a topological nature stemming from the topological defect, while the other is due to the magnetic flux $\Phi$. By adjusting the magnetic flux as an external fine-tuning parameter to counterbalance the topological contribution introduced by the defect, we can nullify the Aharonov–Bohm effect in the ring. In such cases, the energy spectrum resembles that of a particle moving in a quantum ring within a space devoid of any topological defect. This is a very important result. It provides a clue how the effects of a topological defect can be totally or partially compensated by an external filed with regard to the observation of the AB effect. Further consequences on the spin and charge currents are elucidated below. %Now we shall discuss the effect of topological defect in terms of screw dislocation on the electronic spectra. From Eq. (\ref{Egn_Rash_flux}) it is obtained that all the above mentioned properties will remain same except the $m$ values shifted by $k\eta$ amount, i.e., $m \to (m-k\eta)$. The results are depicted in Figs. \ref{fig:EvsR_mag}(c) and (d) considering $\alpha=0.5$ and $\alpha=1$ with the parameter $k\eta=0.3$. Thus for the dice case ($\alpha=1$) the band crossing point occurs at $R=\sqrt{2}l_0\sqrt{\{1/2-(m-k\eta)\}}$ that means a shift in the positive direction, which is shown in Fig. \ref{fig:EvsR_mag}(d). Further, the extrema point conditions in Eqs. (\ref{cond1}) and (\ref{cond2}) will also modified by $m\to (m-k\eta)$, which shift the extrema points by an amount of $k\eta$ towards the positive direction as shown in Figs. \ref{fig:Evsphi}(c)-(f). However the energy gaps at the extrema points (Eq. \ref{gap}) are independent of the topological defect ($k\eta$), only depend on the $\alpha$ and $\lambda_R$ parameters. To show that we have considered two values of $\lambda_R$, namely, $\lambda_R=0.5t$ and $\lambda_R=0.8t$. For $\alpha=1$, the energy gap values are shown in the respective figures.

%Note that in the present case the we have a one-dimensional quantum ring in the presence of a screw dislocation and Aharonov–Bohm flux. The energy spectrum exhibit a parabolic dependence on the burgers vector, similarly to the dependence in the magnetic flux. Analyzing our results we conclude that the particle in a space with a topological defect has behavior like a particle in an Euclidean space in the presence of a effective magnetic flux crossing the ring. This effective flux has two contribution, the first one has topological nature due to topological defect, the other is due to the magnetic flux $\Phi$. We can use the magnetic flux as external fine tuning to compensate the topological contribution introduced by the defect. In this way, no Aharonov–Bohm effect in the ring is observed when the term of magnetic flux compensate the topological contribution, the energy spectrum is similar to that one concerning a particle moving in quantum ring in a space without topological defects.

\begin{figure}[h!]
\centering
\includegraphics[width=9cm, height=10cm]{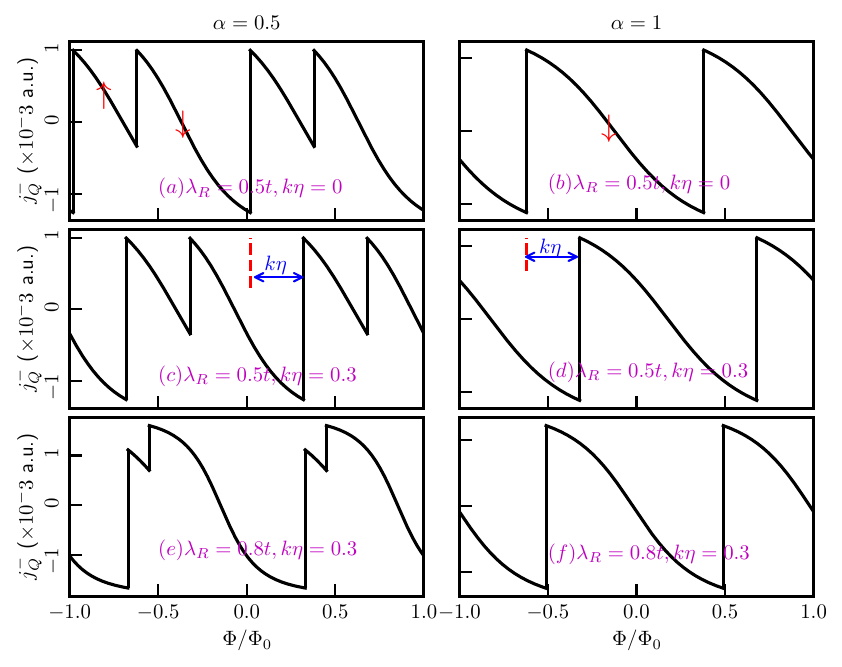}
\caption{(Color online) The charge persistent currents are plotted as functions of the external magnetic flux, considering the low-energy states without accounting for the topological defect, in panels (a) for $\alpha=0.5$ and (b) for $\alpha=1$, with $\lambda_R=0.5t$. In panels (c) and (d), the same currents are presented in the presence of the screw dislocation term $k\eta=0.3$, for $\alpha=0.5$ and $\alpha=1$ respectively. Panels (e) and (f) highlight the impact of the Rashba SOC as we increase the $\lambda_R$ parameter to $0.8t$, while keeping $k\eta=0.3$ fixed for $\alpha=0.5$ and $1$ respectively. In panels (a) and (b), the labels $\uparrow$ and $\downarrow$ denote the currents from the corresponding spin bands. In panels (c) and (d), the phase shift is represented by the blue arrows.}
\label{fig:Jvsphi}
\end{figure}
\subsection{Charge persistent current}
The charge persistent current in the low-energy state can be calculated using the linear response formula, $j_Q=-\sum_{m,\kappa}\frac{\partial E}{\partial \Phi}$, where the sum refers to all (and only) the occupied states (for the valence band ($\kappa=-1$)) and the $m$ values are chosen carefully to perform the summation. Since the current is periodic in $\Phi/\Phi_0$ with a period of 1 (that is $\Phi=\Phi_0$), we restrict the discussion to the window $-1\leq \Phi/\Phi_0 \leq 1$. The analytical form for the charge current is obtained as,
%\begin{widetext}
\begin{equation}
\label{per_curr_rash}
\begin{aligned}
j_{Q}^\kappa &=-\frac{\epsilon^2\kappa}{2\Phi_0}\sum_m\frac{\big(1+\frac{\lambda_R^2}{t^2}\big)\Big[2\big(m-k\eta+\frac{\Phi}{\Phi_0}\big)-\frac{1-\alpha^2}{1+\alpha^2}\Big]}{E_2(\Phi)}\\
&-\frac{\epsilon^2\kappa}{2\Phi_0}\sum_m\frac{2\big(m-k\eta+\frac{\Phi}{\Phi_0}\big)\big(1+\frac{\lambda_R^2}{t^2}\frac{1-\alpha^2}{1+\alpha^2}\big)+\big(\frac{\lambda_R^2}{t^2}+\frac{1-\alpha^2}{1+\alpha^2}\big)}{E_3(\Phi)}.
\end{aligned}
\end{equation}
%\end{widetext}

The spin branches closest to the Fermi energy exhibit non-monotonic behaviour, resulting in two distinct contributions to the charge current coming from the $\uparrow$-spin and $\downarrow$-spin components. Since we are calculating the current contributions arising from the low-energy states, it is clear from Fig. \ref{fig:Evsphi} that for a certain range of $\Phi/\Phi_0$, only one energy state labelled by a particular value of $m$ is present. Hence, the sum in Eq. (\ref{per_curr_rash}) comprises of only one value of $m$. Furthermore, based on the observations in Fig. \ref{fig:Evsphi}, we can discern that the low-energy states when $\alpha = 0.5$ encompass both the spin bands. Consequently, in our current calculations, we considered contributions from both the spin branches. In contrast, for the $\alpha = 1$ scenario, the low-energy state exclusively comprises the $\downarrow$-spin branches, and thus, we only have accounted for the contributions from the $\downarrow$-spin to compute the persistent current. The outcomes of these calculations are depicted in Fig. \ref{fig:Jvsphi} for both $\alpha = 0.5$ and $\alpha = 1$, considering various combinations of $\lambda_R$ and $k\eta$, all the while maintaining a fixed ring radius of $R = 10$ $nm$. The asymmetric spectral features between the two spin branches allows for the possibility of a net spin currents, as we shall see below. For all values of $\alpha$, the persistent currents oscillate periodically with $\Phi/\Phi_0$, with a periodicity of $\Phi/\Phi_0=1$. Figs. \ref{fig:Jvsphi}(a) and (b) illustrate that the persistent currents can be tuned by adjusting the parameter $\alpha$ for a fixed value of the Rashba coupling ($\lambda_R$). Moreover, the charge persistent currents can be manipulated via $\lambda_R$ for a fixed $\alpha$, (see Fig. \ref{fig:Jvsphi}(a), (c) and (e)) since the Rashba parameter can be controlled by a gate voltage. Further, it is worth noting the presence of a kink in the current profile when $\alpha < 1$. This kink arises because different spin bands contribute to the current, as indicated by the distinctions denoted by $\uparrow$ and $\downarrow$ for their respective spin bands. However, this kink phenomenon is absent for the dice lattice ($\alpha = 1$), as evident in Figure \ref{fig:Jvsphi}(b), (d), and (f). The reason being, in this case, only the $\downarrow$-spin bands contribute to the current, as mentioned earlier.

Furthermore, the topological defect plays a pivotal role in the behaviour of the charge current. As illustrated in Figure \ref{fig:Jvsphi}, we can observe that, for a fixed value of $\lambda_R$, the screw dislocation induces a phase shift in the current, regardless of the specific value of $\alpha$. For example, every point of the current profile is shifted by the same amount as the strength of the topological defect as shown in the middle panel of Fig. \ref{fig:Jvsphi}. However, the topological defect does not influence the current profile itself. A desired shift in the current profile may be achieved via the controllable parameter $k\eta$. Further, the Rashba coupling, $\lambda_R$ plays a role as well. The depth in the kink of the current profile decreases with the increasing $\lambda_R$ (see bottom panel of Fig. \ref{fig:Jvsphi}). This can be understood from the lower panel of Figure \ref{fig:Evsphi}, as the increase in $\lambda_R$ results in a decrease in the $\uparrow$-spin contribution in the low-energy state within a certain range of $\Phi/\Phi_0$. However, the overall oscillation period remains unaltered.

In summary, we can manipulate the persistent current profile by fine-tuning the Rashba coupling, and we have the ability to shift the phase of the current to suit our specific applications by adjusting the strength of the topological defect.

\subsection{Equilibrium spin currents}
\begin{figure}[h!]
\centering
\includegraphics[width=9cm, height=10cm]{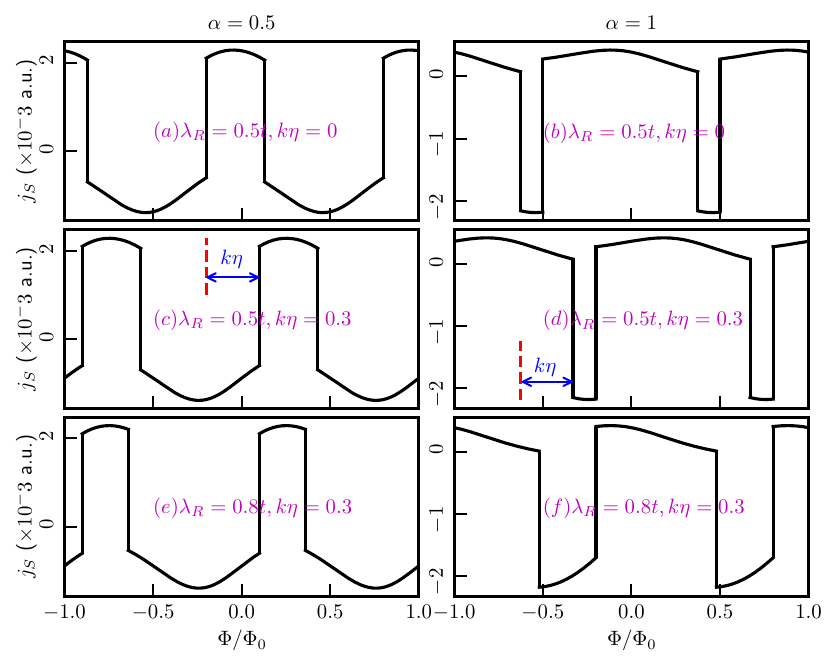}
\caption{(Color online) The equilibrium spin currents are plotted as functions of the external magnetic flux, considering the low-energy states without accounting for the topological defect, in panels (a) for $\alpha=0.5$ and (b) for $\alpha=1$, with $\lambda_R=0.5t$. In panels (c) and (d), the same currents are presented in the presence of the screw dislocation term $k\eta=0.3$, for $\alpha=0.5$ and $\alpha=1$ respectively. Panels (e) and (f) highlight the impact of the Rashba SOC as we increase the $\lambda_R$ parameter to $0.8t$, while keeping $k\eta=0.3$ fixed for $\alpha=0.5$ and $1$ respectively. In panels (c) and (d), the phase shift is represented by the blue arrows.}
\label{fig:Spin_cur}
\end{figure}

We shall now study equilibrium spin currents. In contrast to the formalism for obtaining the charge current, one can obtain the spin currents by accounting for distinct velocities for different spin branches. Thus, we define equilibrium spin current as,
\begin{equation}
j_S=j_Q(\uparrow)-j_Q(\downarrow).
\label{Eq.spin_cur}
\end{equation}
We have calculated the equilibrium spin currents following the procedure discussed earlier. The peculiar separation of the spin branches results in differences of the velocities between the two spin projections, giving rise to a spin current, as shown in Fig. \ref{fig:Spin_cur}. The figure illustrates a significant spin current for small values of the flux, which can be attributed to the large charge current originating from a single spin branch.

The striking feature is that the magnitude as well as the pattern of the spin currents depend upon the parameters $\alpha$ and the strength of the Rashba coupling ($\lambda_R$). We present results for $\alpha=0.5$, and $\alpha=1$ with different values of $\lambda_R$ and topological defect $k\eta$ for a ring of radius, $R=10$ $nm$. The presence of the Rashba coupling breaks inversion symmetry (in addition to the $\sigma_z$ symmetry) in the plane even for small $\lambda_R$. The symmetry breaking determines the spin labelling of the energy branches that take part in yielding the spin currents. Additionally, the spin currents exhibit periodic behaviour with $\Phi/\Phi_0$, with a periodicity equal to one flux quantum, irrespective of $\alpha$. Similar to the case of charge persistent current, we also notice a phase shift introduced by the topological defect. The magnitude of this phase shift precisely corresponds to the strength of the defect.

These findings underscore the potential of $\alpha$-$T_3$ quantum rings as key components for spintronic applications, where we can manipulate the performance of the devices by adjusting both the Rashba coupling and the topological defect.

\section{Thermodynamic properties of the $\alpha$-$T_3$ AB ring}
\label{Thermodyna}
\subsection{Formalism}
Having studied the transport features, it may be of interest to explore the thermodynamic properties such as, specific heat, bulk modulus, compressibility, etc. Specifically, to the best of our knowledge these quantities have never been explored in the presence of Rashba coupling and topological defects in the context of a QR. However, we shall restrict ourselves in studying the thermodynamic potentials, entropy, internal energy, and the specific heat in the following. 

In this section, we will study the thermodynamic properties of an AB $\alpha$-$T_3$ ring with Rashba coupling, which is in contact with a thermal reservoir at a finite temperature. These properties encompass fundamental thermodynamic quantities, specifically the Helmholtz free energy, the internal energy, entropy, and heat capacity. We shall exclusively focus on stationary states characterized by positive energies ($E > 0$) that constitute our thermodynamic ensemble. Given the strict exclusion of particle-particle interactions, the excitation of negative-energy states and the issue of pair production are excluded for the discussions [\onlinecite{Gran,Rach}]. Consequently, the partition function involves a summation solely over positive-energy states, simplifying the analysis considerably. We commence by assessing the corresponding partition function as,
\begin{equation}
\label{Partition}
Z=\sum_{m=0}^\infty(e^{-\beta E_2^+}+e^{-\beta E_3^+})
\end{equation}
where $\beta=\frac{1}{k_BT}$, $k_B$ is the Boltzmann constant and $T$ is the equilibrium temperature. Rearranging the expressions in Eq. (\ref{Egn_Rash_flux}) and using Eq. (\ref{Partition}), we obtain that for a one-fermion confined in the $\alpha$-$T_3$ AB ring the partition function is,
\begin{equation}
\label{Partition1}
Z=\sum_{m=0}^\infty e^{-\beta\sqrt{A_1m^2+B_1m+C_1}}+\sum_{m=0}^\infty e^{-\beta\sqrt{A_2m^2+B_2m+C_2}}
\end{equation}
where the quantities are,
\begin{equation}
\begin{aligned}
A_1 &=\epsilon^2\Big(1+\frac{\lambda_R^2}{t^2}\Big)\\
B_1 &=\epsilon^2\Big(2\frac{\Phi}{\Phi_0}-2k\eta-\cos2\xi\Big)\Big(1+\frac{\lambda_R^2}{t^2}\Big)\\
C_1 &=\epsilon^2\Big\{\frac{1}{4}+\big(k\eta-\frac{\Phi}{\Phi_0}\big)^2+\big(k\eta-\frac{\Phi}{\Phi_0}\big)\cos2\xi \Big\}\Big(1+\frac{\lambda_R^2}{t^2}\Big)\\
A_2 &=\epsilon^2\Big(1+\frac{\lambda_R^2}{t^2}\cos2\xi\Big)\\
B_2 &=\epsilon^2\Big\{2\big(\frac{\Phi}{\Phi_0}-k\eta\big)\Big(1+\frac{\lambda_R^2}{t^2}\cos2\xi\Big)+\Big(\frac{\lambda_R^2}{t^2}+\cos2\xi\Big)\Big\}\\
C_2 &=\epsilon^2\Big[\Big\{\frac{1}{4}+\big(k\eta-\frac{\Phi}{\Phi_0}\big)^2\Big\}\Big(1+\frac{\lambda_R^2}{t^2}\cos2\xi\Big)+\big(\frac{\Phi}{\Phi_0}-k\eta\big)\\
&\Big(\frac{\lambda_R^2}{t^2}+\cos2\xi\Big)\Big].
\end{aligned}
\end{equation}
Since Eq. (\ref{Partition1}) cannot be computed in a closed form, we assume that the AB ring is submitted to a strong magnetic field ($\Phi\gg \Phi_0$). Therefore, the expression under the summation takes the approximate form $e^{-\beta\sqrt{B_{1/2}m+C_{1/2}}}$, which is a monotonically decreasing function and the associated integral,
\begin{equation}
\label{approx}
\int_0^\infty e^{-\beta\sqrt{Bx+C}}dx=\frac{2}{B\beta^2}(1+\beta\sqrt{C})e^{-\beta\sqrt{C}},
\end{equation}
is finite. Thus, the form ensures the convergence of the series.
To evaluate the partition function (\ref{Partition1}) at the strong-field limit we will use the Euler-Maclaurin sum formula given by [\onlinecite{Grei,Oliv}],
\begin{equation}
\label{Euler}
\sum_{n=0}^\infty f(n)= \frac{1}{2}f(0)+\int_0^\infty f(x)dx - \sum_{p=1}^\infty \frac{1}{(2p)!}B_{2P}f^{(2p-1)}(0),
\end{equation}
where $B_{2p}$ are the Bernoulli numbers, $B_2=1/6$, $B_4=-1/30$,.... They are defined through the series
\begin{equation}
\frac{x}{e^x-1}=\sum_{n=0}^\infty B_n\frac{x^n}{n!}.
\end{equation}
Therefore, Eq. (\ref{Euler}) can be rewritten simply as,
\begin{equation}
\label{Simple}
\sum_{n=0}^\infty f(n)\simeq \frac{1}{2}f(0)+\int_0^\infty f(x)dx -\frac{1}{12}f'(0)+\frac{1}{720}f'''(0)-...,
\end{equation}
According to the expression above, we can write the partition function (\ref{Partition1}) explicitly as,
\begin{widetext}
\begin{equation}
\label{Partition2}
\begin{split}
Z\simeq e^{-\beta\sqrt{C_1}}\Big[\frac{2}{B_1\beta^2}\Big(1+\beta\sqrt{C_1}\Big)+\frac{1}{2}+\Big(\frac{B_1}{24\sqrt{C_1}}-\frac{B_1^3}{720\sqrt{C_1^5}}\Big)\beta+\frac{1}{90}\Big(\frac{A_1}{2C_1}-\frac{B_1^2}{8C_1^2}\Big)\beta^2-\mathcal{O}(\beta^3) \Big]\\
 +e^{-\beta\sqrt{C_2}}\Big[\frac{2}{B_2\beta^2}\Big(1+\beta\sqrt{C_2}\Big)+\frac{1}{2}+\Big(\frac{B_2}{24\sqrt{C_2}}-\frac{B_2^3}{720\sqrt{C_2^5}}\Big)\beta+\frac{1}{90}\Big(\frac{A_2}{2C_2}-\frac{B_2^2}{8C_2^2}\Big)\beta^2-\mathcal{O}(\beta^3) \Big]
\end{split}
\end{equation}
\end{widetext}
where $f(n)=e^{-\beta\sqrt{B_{1/2}n+C_{1/2}}}$ and $\mathcal{O}(\beta^3)$ represents terms involving high order in $\beta$ which will be neglected henceforth.
Notice that when one considers the high-temperatures regime ($\beta \ll 1$), Eq. (\ref{Partition2}) becomes,
\begin{equation}
\label{Partition3}
Z\simeq \frac{2}{B_1\beta^2}\Big(1+\beta\sqrt{C_1}\Big)+\frac{2}{B_2\beta^2}\Big(1+\beta\sqrt{C_2}\Big).
\end{equation}
The stability of graphene at high temperature [\onlinecite{Kim}], provides motivation to study the thermodynamic properties the $\alpha$-$T_3$ AB ring. Indeed, the total partition function $Z_N$ for $N$ fermions can be approximated by,
\begin{equation}
\label{Partitin4}
Z_N\simeq \Bigg[\frac{2}{B_1\beta^2}\Big(1+\beta\sqrt{C_1}\Big)+\frac{2}{B_2\beta^2}\Big(1+\beta\sqrt{C_2}\Big)\Bigg]^N.
\end{equation}
Therefore, using the partition function in Eq. (\ref{Partitin4}), we can now determine all related thermodynamic quantities. Indeed, after some algebra we obtain the Helmholtz free energy $F$, the internal energy $U$, the entropy $S$ and the heat capacity $C_V$ as follows:
\begin{widetext}
\begin{equation}
\label{Helm}
%\begin{split}
F =-\frac{1}{\beta}\mathrm{ln}Z_N = -\frac{N}{\beta}\mathrm{ln}\Bigg[\frac{2}{B_1\beta^2}\Big(1+\beta\sqrt{C_1}\Big)+\frac{2}{B_2\beta^2}\Big(1+\beta\sqrt{C_2}\Big)\Bigg].
%\end{split}
\end{equation}
\begin{equation}
\label{Internal}
%\begin{split}
U =-\frac{\partial}{\partial\beta}\mathrm{ln}Z_N = N\frac{B_2\Big(2+\beta\sqrt{C_1}\Big)+B_1\Big(2+\beta\sqrt{C_2}\Big)}{\beta\Big[B_2\Big(1+\beta\sqrt{C_1}\Big)+B_1\Big(1+\beta\sqrt{C_2}\Big)\Big]}.
%\end{split}
\end{equation}
\begin{equation}
\label{Entropy}
%\begin{split}
S =k_B\beta^2\frac{\partial F}{\partial\beta} = Nk_B\Bigg[ \mathrm{ln}\Big\{\frac{2\Big(1+\beta\sqrt{C_1}\Big)}{B_1\beta^2}+\frac{2\Big(1+\beta\sqrt{C_2}\Big)}{B_2\beta^2}\Big\}+\frac{B_2\Big(2+\beta\sqrt{C_1}\Big)+B_1\Big(2+\beta\sqrt{C_2}\Big)}{B_2\Big(1+\beta\sqrt{C_1}\Big)+B_1\Big(1+\beta\sqrt{C_2}\Big)} \Bigg].
%\end{split}
\end{equation}
\begin{equation}
\label{Heat}
%\begin{split}
C_V =-k_B\beta^2\frac{\partial U}{\partial\beta} = Nk_B\Bigg[ \frac{2\Big\{B_2\Big(3+\beta\sqrt{C_1}\Big)+B_1\Big(3+\beta\sqrt{C_2}\Big)\Big\}}{\Big\{B_2\Big(1+\beta\sqrt{C_1}\Big)+B_1\Big(1+\beta\sqrt{C_2}\Big)\Big\}}-\frac{\Big\{B_2\Big(2+\beta\sqrt{C_1}\Big)+B_1\Big(2+\beta\sqrt{C_2}\Big)\Big\}^2}{\Big\{B_2\Big(1+\beta\sqrt{C_1}\Big)+B_1\Big(1+\beta\sqrt{C_2}\Big)\Big\}^2} \Bigg].
%\end{split}
\end{equation}
\end{widetext}
%\begin{widetext}

\begin{figure}[h!]
\centering
\includegraphics[width=9cm, height=9cm]{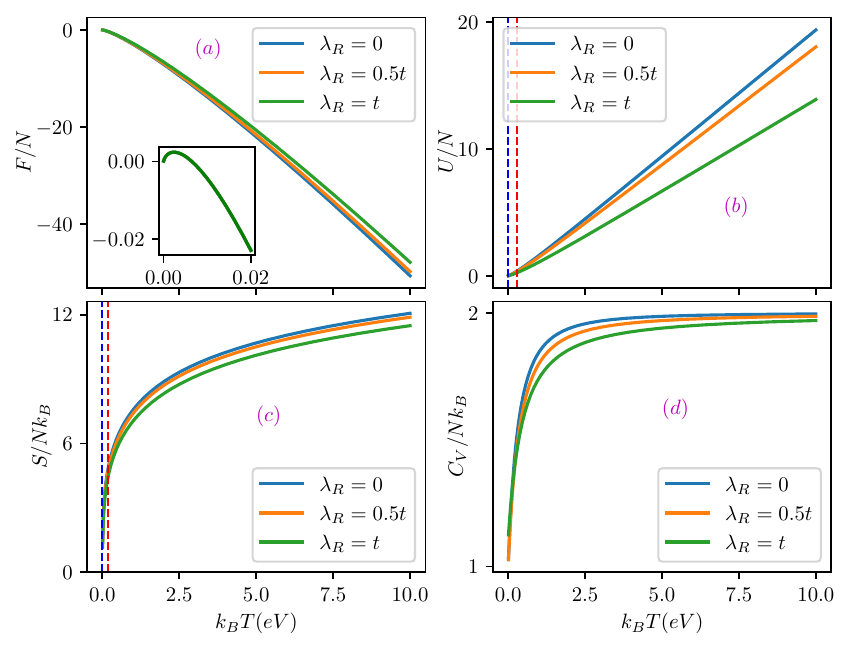}
\caption{(Color online) Thermodynamic quantities of the Aharonov-Bohm $\alpha$-$T_3$ ring are displayed as functions of temperature $k_BT$, with varying values of $\lambda_R = 0, 0.5t$, and $t$. These results are obtained for a fixed magnetic flux of $\Phi=20\Phi_0$. (a) the Helmholtz free energy, (b) the internal energy, (c) the entropy, and (d) the heat capacity. The ring radius is set to $R=10$ nm, and the parameter $k\eta$ is fixed at $0.3$.}
\label{fig:Thermo}
\end{figure}
%\end{widetext}

\subsection{Results and discussions}
%In the sequel, we used the Euler-Maclaurin formula regarding a strong manetic field to evaluate the partition function. We recall our theory includes two interesting quantities, which are the Rashba coupling parameter $\lambda_R$ and the external magnetic flux $\Phi$. In Fig. {\ref{fig:Thermo} we investigate all the profiles of the thermal quantites given in Eqs. (\ref{Helm})-(\ref{Heat}) as a function of temperature $k_BT$ by shooting the values of the Rashba coupling parameters, $\lambda_R=0,0.5t$ and $t$ for a fixed value of $\Phi=20\Phi_0$. From Fig. \ref{fig:Thermo}(a) we observe that the Helmholtz function $\frac{F}{N}$ decreases with a nearly linear behaviour when $k_BT$ increases and has higher values when $\lambda_R$ increases. In Fig. \ref{fig:Thermo}(b) for the very low temperature we observe that the behaviour of the internal energy $\frac{U}{N}$ is independent on the different values of $\lambda_R$. While, after the critical temperature it starts increasing with a nearly linear behaviour. As $\lambda_R$ decreases, the curve of the internal energy grows faster. The entropy $\frac{S}{Nk_B}$ is independent of the Rashba coupling in the interval $0<k_BT<0.2 eV$, but when $k_BT>0.2 eV$, $\frac{S}{Nk_B}$ shows a small variation in terms of $\lambda_R$, as decreases for large $\lambda_R$'s as depicted in Fig. \ref{fig:Thermo}(c). The heat capacity $\frac{C}{Nk_B}$ tends to an asymptotic behaviour fixed in the value $2$ when $k_BT$ increases.

Subsequently, we applied the Euler-Maclaurin formula, especially tailored for a strong magnetic field, to compute the partition function. It is worth noting that our theory involves two key parameters of interest, the Rashba coupling parameter, $\lambda_R$, and the external magnetic flux, $\Phi$. The third participant, namely, the topological defect is a indispensable quantity, owing to its ability to lift spectral degeneracies, however, the thermodynamic quantities remain insensitive to its magnitude. We have done calculations for a fixed value of dislocation, that is, $k\eta=0.5$ and have considered only $\alpha=0.5$. The other value of $\alpha$, namely $\alpha=1$ does not inflict any significant changes. In Fig. {\ref{fig:Thermo}, we investigate the profiles of various thermal quantities outlined in Eqs. (\ref{Helm})-(\ref{Heat}) as a function of temperature ($k_BT$), while varying the values of the Rashba coupling parameter, $\lambda_R$, with various values of $\lambda_R = 0, 0.5t,$ and $t,$ while keeping the external magnetic flux constant at $\Phi = 20\Phi_0$ (necessity of such a large field is elaborated earlier).

In Fig. \ref{fig:Thermo}(a), we observe that the Helmholtz function, $F/N$, exhibits a small enhancement when $k_BT$ increases at very low temperatures which is shown in the inset figure. However, following a small hump, it has a nearly linear decrease as $k_BT$ increases, it becoming less steeper at larger values of $\lambda_R$. In Fig. \ref{fig:Thermo}(b), for low temperatures, $0<k_BT<0.3$ $eV$, the behaviour of the internal energy, $U/N$, remains unaffected by $\lambda_R$. However, after reaching a certain critical temperature ($k_BT\simeq 0.3$ $eV$), it begins to increase almost linearly. The growth of the internal energy is more rapid (steeper) with decreasing $\lambda_R$. The entropy, $S/Nk_B$, is independent of the Rashba coupling within the range of $0 < k_BT < 0.2$ $eV$, however, beyond this temperature threshold ($k_BT > 0.2$ $eV$), $S/Nk_B$ nearly exponential increase with a slight variations with respect to $\lambda_R$. Notably, $S/Nk_B$ decreases for larger values of $\lambda_R$, as illustrated in Fig. \ref{fig:Thermo}(c). The heat capacity, $C/Nk_B$, rises quickly at small $T$, and tends to an asymptotic behaviour, converging to a fixed value of $2$ as $k_BT$ increases which is shown in Fig. \ref{fig:Thermo}(d). This is an expected result for a 1D Dirac oscillator attached with a thermal bath [\onlinecite{Pach,Boum}] and is reproduced by us for a QR. We have further checked the scenario corresponding to different values of the magnetic flux. The quantitative behaviour of the thermodynamic quantities remain unaltered as those with varying the spin-orbit coupling term.

By analysing the profiles of the main thermodynamic functions, we can gain insights into the thermal behaviour of our system. We observe that the Helmholtz free energy exhibits a non-monotonic behaviour at low temperatures. As the temperature increases, the available free energy for performing ensemble work rapidly diminishes as seen from Fig. \ref{fig:Thermo}(a). Conversely, when this ensemble achieves thermal equilibrium within the reservoir, the system's internal energy steadily increases. Additionally, the thermal variation of the $\alpha$-$T_3$ AB ring tends to rise until it reaches thermal equilibrium in both scenarios. Once equilibrium is achieved, the heat capacity remains constant. Moreover, in accordance with principles from solid-state physics, we can observe that our system satisfies the well-known Dulong-Petit law for a Dirac oscillator in 1D. This law is characterized by $C_V/N\simeq 2k_B$, as evident in Fig. \ref{fig:Thermo}(d). It is worth noting that the presence of screw dislocations (topological defect) does not play an actice role in the high-flux limit. However, its presence is necessary to remove degeneracies and enables usage of the Eq. (\ref{approx}). Furthermore, the qualitative features of the thermodynamic functions remain consistent across all values of $\alpha$ in the range [0:1].

\section{Summary and conclusions}
\label{Sum}
In summary, we have conducted a comprehensive investigation of the properties of Rashba spin-orbit coupling in the context of an $\alpha$-$T_3$ pseudospin-1 fermionic Aharonov-Bohm quantum ring, considering the presence of a special type of topological defect. Our exploration covered aspects such as the energy spectrum, persistent currents, their dependencies on spin-orbit couplings, screw dislocations, and magnetic fields. Additionally, we delved into the thermodynamic characteristics of the ring in contact with a heat bath. Our key findings are as follows.

The introduction of the spin-orbit coupling parameter, $\lambda_R$, results in a spectrum composed of six bands, including two non-dispersive flat bands one for each spin, and four dispersive spin-split valence and conduction bands. The flat bands encompass a multitude of degenerate levels at zero-energy, which remain unaffected by applied magnetic fields. In the absence of a magnetic field, the energy levels in both the conduction and valence bands exhibit inverse dependence on the ring radius, denoted as $R$, and are independent of $\alpha$. Notably, $\uparrow$-spin energy levels are two-fold degenerate for $\alpha=1$, except for the $m=0$ level, while the $\downarrow$-spin bands are non-degenerate for all $\alpha$ values. The presence of a screw dislocation, serving as a topological defect, renders splitting of degeneracy, although other features remain unaltered. Under the influence of a perpendicular magnetic field, the energy levels deviate significantly from their usual $R$-dependence, showcasing behaviours of $\sim 1/R$ for small $R$ and $\sim R$ for larger $R$. The presence of the topological defect introduces a topological term, effectively acting as a magnetic flux traversing the ring. This effective flux is the result of two contributions, one stemming from the topological nature of the defect and the other from the magnetic flux.

Furthermore, the persistent currents in both the spin and charge sectors exhibit periodic oscillations with a periodicity of $\Phi_0$, featuring distinct patterns corresponding to different $\alpha$ and $\lambda_R$ values. Notably, the presence of a topological defect shifts the phase of current oscillation by an amount equal to the strength of the defect. We have also derived equilibrium spin currents by combining the charge current contributions from different spin branches, underscoring the potential utility of our system in spintronic applications. Equilibrium spin currents are present for all $\alpha$ values, with a $\Phi=\Phi_0$ periodic behaviour. Similar to the charge persistent current, the presence of a topological defect shifts the phase of the oscillations in the spin current profile by an amount proportional to the defect.

Further, to address the thermodynamic properties, we have computed the canonical partition function. As the exact partition function lacks a closed form, we resorted to the strong field approximation and utilized the Euler-Maclaurin summation formula for its numerical evaluation. Once the partition function was determined, it paved the way for deriving all the key thermodynamic quantities, including the Helmholtz free energy $F$, internal energy $U$, entropy $S$, and heat capacity $C_V$. We generated graphical representations of these quantities for various $\lambda_R$ values in presence of a topological defect. The outcomes of our study highlighted emergence of the well-established Dulong-Petit law for the specific heat. We anticipate that our findings will serve as a valuable tool for examining these thermal properties in connection with experimental investigations.

In conclusion, by adjusting the parameters $\alpha$, $k\eta$, and $\lambda_R$, we have comprehensively shown the ability to manipulate the persistent currents, rendering them as controllable features in our system.

\section*{ACKNOWLEDGMENTS}
M. I. sincerely acknowledges the financial support from the Council of Scientific and Industrial Research (CSIR), Govt. of India  to pursue this work.

\end{document}